\documentclass[twocolumn,showpacs,preprintnumbers,amsmath,amssymb]{revtex4}

\usepackage{graphicx}
\usepackage{dcolumn}
\usepackage{bm}
\usepackage{amssymb}

\newcommand\vdecay{V \rightarrow e^+  e^-}
\newcommand\xdecay{X \rightarrow e^+  e^-}
\newcommand\brx{Br(\pi^0 \to \gamma X)}

\newcommand\pair{e^+ e^-}
\newcommand\ee{e^+e^-}

\def\address{\@ifstar{\address@star}%
  {\@ifnextchar[{\address@optarg}{\address@noptarg}}}

\begin{document}

\author{S.N.~Gninenko}
\affiliation{Institute for Nuclear Research, Moscow 117312}


\title{Stringent limits on the $\pi^0 \to \gamma X,~\xdecay$ decay from neutrino experiments and constraints on new light gauge bosons}

\date{\today}

\begin{abstract}
We report new limits on the $\pi^0 \to \gamma X$  decay of 
the neutral pion into a photon and a new gauge boson $X$ followed by the decay $\xdecay$. If this 
process exist, one would expect a flux of high energy $X$'s produced from $\pi^0$'s  generated by the proton beam in a neutrino target. The $X$'s would then penetrate the downstream shielding and be observed in a neutrino detector via their decays. Using bounds from the NOMAD and PS191 neutrino experiments at CERN that searched for an excess of $\ee$ pairs from heavy neutrino decays, stringent limits  on the branching ratio as small as $Br(\pi^{0}\to \gamma X) \lesssim 10^{-15}$ are  obtained. These limits are several orders of magnitude smaller than the previous experimental and cosmological bounds.  The obtained results are used to constrain models, where the $X$ interacts with quarks and 
 leptons, or it is a new vector boson mixing with 
  photons, that transmits    interaction between our world and  
 hidden sectors consisting of  $SU(3)_C \times SU(2)_L \times U(1)_Y$ singlet fields. 
\end{abstract}
\maketitle
Many extensions of the Standard Model (SM)  such as  GUTs~\cite{1}, 
super-symmetric~\cite{2}, super-string models~\cite{3} and models including a
new long-range interaction, i.e. the fifth force \cite{carl}, predict an extra
U$^{'}$(1) factor and therefore the existence of a new gauge boson $X$
corresponding to this new group.
The predictions for the  mass of the  $X$ boson are not very firm and it could
be light enough ($M_{X}\ll M_{Z}$) for searches  at low energies.
If the mass $M_X$ is  of the order of the pion mass,
 an effective search could be conducted for this new vector boson in the radiative decays of neutral  pseudoscalar mesons $P\rightarrow\gamma X$, where $P = \pi^{0},\eta$, or 
$\eta^{\prime}$, because  the decay rate of  $P\rightarrow\gamma~+~$ 
$\it any~new~particles~with~spin~0~or~\frac{1}{2}$ proves to be negligibly 
small~\cite{di}.\
Therefore, a positive result in the direct search for these decay modes could be
interpreted unambiguously as the discovery of a new light spin 1 particle, in
contrast with other experiments searching for light weakly interacting particles
in rare K, $\pi$ or $\mu$ decays~\cite{di, md, gkx2}.
 
The decay $\pi^0 \to \gamma X$ can  also occur in several interesting extensions of the SM  that suggest the existence of, so-called, ``hidden'' sectors consisting of $SU(3)_C \times SU(2)_L \times U(1)_Y$ singlet fields. These  are sectors of  weakly interacting massive particles (WIMPs), that do not interact with the ordinary matter directly and  couple to it by gravity and possibly by other very weak forces.  If the mass scale of a hidden sector is too high, it  will be  experimentally unobservable. However, there is  a class of models where Dark Matter WIMPs have  interactions with the SM mediated by 
U$_\mathrm{h}(1)$ gauge group and the corresponding hidden gauge boson could be light, see e.g. \cite{jr, rw, prv}.
The fact that the $X$ coupling strength may be in the experimentally accessible  and theoretically interesting regions, makes  further searches for $X$'s   very attractive at the high intensity frontier \cite{hif}.

From the analysis of data from earlier experiments, constraints on the
branching ratio for the decay of  $P\rightarrow\gamma + X$ range from 
$10^{-7}$ to $10^{-3}$ depending on whether $X$ interacts with both quarks and
leptons or only with quarks \cite{pdg}.
Direct searches for a signal from $\pi^{0}\rightarrow\gamma X$ decay have been
performed in a few experiments with two different methods: i) searching for a
peak in inclusive photon spectra from two-body 
$\pi^{0}\rightarrow\gamma+nothing$ decays, where ``nothing'' means that $X$ is
not detected because it either has a long life time or decays into
$\nu\overline{\nu}$ pairs \cite{mei1, mei2, ati, amsl1, amsl2}, and 
ii) searching  for a peak in the invariant mass spectrum of $e^{+}e^{-}$ pairs
from $\pi^{0}$ decays, which corresponds to the decay 
$X\rightarrow e^{+}e^{-}$~\cite{mei1, mei2}.

The best experimental limit on the branchning ratio 
of the decay $\pi^{0}\rightarrow\gamma X$, $\brx < (3.3 - 1.9)\times 10^{-5}$  ($90\% C.L.$) 
for $X$ masses ranging from 0 to 120 MeV, was obtained by the NOMAD experiment at CERN \cite{nomadx}.
Using 450 GeV proton collisions with the SPS neutrino target as  a source of high energy  X's from $\pi^{0}\rightarrow \gamma X$ decays, they searched for a signal from the Primakoff conversion $X\to \pi^0$ in their detector \cite{gkx1}. 
The best bound for the branching ratio 
$Br(\pi^0\to \gamma X) Br(X\to \ee) < 2\times 10^{-4} - 4\times 10^{-6}$ 
for the $X$ mass range $25<M_{X}<120~MeV$ was obtained by the SINDRUM collaboration at 
PSI \cite{mei1}. 
In this letter we show that  more stringent limits on the decay $\pi^{0}\rightarrow\gamma X$, followed by the decay $\xdecay$ can be 
obtained  from the results of sensitive searches for an excess of single
isolated $\ee$ pairs from decays of heavy neutrinos in the sub-GeV  mass range   by the  PS191 \cite{ps191-1, ps191-2} and NOMAD \cite{nomadnuh, nomadkarm, nomadsg} experiments at CERN.

Consider first the NOMAD experiment on search for the  decay $\nu_h\to \nu e^+ e^-$ 
of a heavy neutrino $\nu_h$ into a lighter neutrino and $\ee$ pair  performed 
at the  CERN West Area Neutrino Facility (WANF) \cite{nomadnuh}.
The WANF provides an essentially pure $\nu_{\mu}$
beam for neutrino experiments. It consists of a Be
target irradiated by  450 GeV  protons from the CERN SPS.
The secondary hadrons are focused by the horn and 
the reflector, and  protons that have not interacted in the target, secondary hadrons and 
muons that do not decay are absorbed by a 400 m thick shielding made of iron and 
earth.  The NOMAD detector  described in Ref.~\cite{nomad} is located at 835 m from the neutrino target. It consists of a number of sub-detectors most of which are located inside a 0.4 T dipole 
magnet  with a  volume of $3.5\times 3.5\times 7.5$ m$^{3}$ including  
 an active target of 
drift chambers (DC) with a mass of 2.7 tons.
 \begin{figure}[tbh!]
\includegraphics[width=0.45\textwidth]{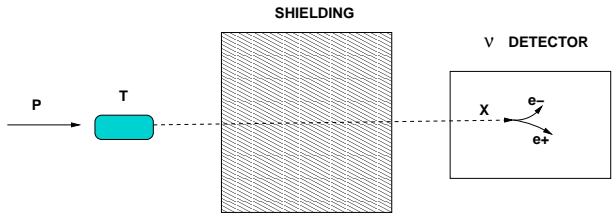}
\caption{ Schematic illustration of a proton beam dump  experiment on search for 
$\pi^0 \to \gamma X \to \ee$ decay chain: neutral pions generated by the proton beam in the 
  neutrino target (T) produce a flux of high energy $X$'s which penetrate the 
 downstream shielding and decay into $\ee$ pair in a neutrino detector.}
\label{setup}
\end{figure} 

 If the  decay $\pi^{0}\to \gamma X$ exists, one expects a  flux of 
high energy $X$ bosons from the WANF target, since $\pi^{0}$'s are
abundantly produced in the forward direction  by high energy  protons either 
in the target or in the beam dump following the decay tunnel.
If $X$ is a relatively long-lived particle, this flux  would penetrate the downstream 
shielding without significant attenuation  and would be observed in the NOMAD 
detector via the $\xdecay$ decay into a high energy $\ee$ pair, as schematically 
illustrated in Fig. \ref{setup}.
The occurrence of $\xdecay$ decays  would appear as an excess
of isolated $\pair$ pairs in NOMAD above those expected from  standard neutrino
interactions. The experimental signature of these events is clean and they can be selected with small background due to the excellent NOMAD capability for precise measurements of $\ee$ pairs, 
see example \cite{nomadpi0, nomadsg}. As  the final states of the 
 decays $\nu_h\to \nu \ee$ and $\xdecay$ are identical, the  
results of the  searches for the former  can be used  
to constrain the later for the same  $\ee$  invariant mass regions.
 
\begin{figure}
\includegraphics[width=0.45\textwidth]{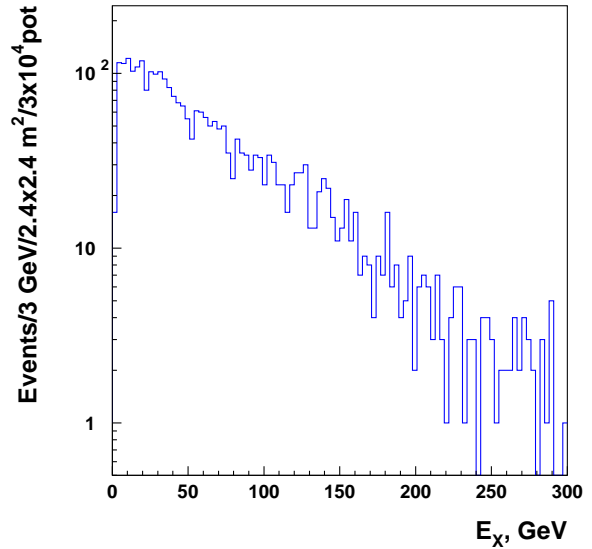}
\caption{ Combined energy spectrum of X bosons with mass M$_{X}$=10 MeV from the 
     SPS neutrino target and from the beam dump region at the NOMAD detector calculated 
     for $\brx = 1$.}
\label{spectr}
\end{figure}

The calculated flux and energy spectrum of $\pi^0$ produced 
in the Be target were used to predict the  
flux of $X$'s with momenta pointing to the NOMAD fiducial area as a  function of the $X$ mass, see Fig. \ref{spectr}. 
The search for the $\nu_h\to \nu \ee$ decay described in \cite{nomadnuh}  corresponds to the total  
number of  $4.1\times 10^{19}$ protons on target (pot).  
The strategy of the analysis was to identify $\nu_h\to \nu \ee$ candidates 
by reconstructing in the DC isolated low invariant mass $\pair$ pairs  
 that are accompanied by no other activity in the detector. The measured rate 
of $\pair$ pairs was then compared to that expected from known sources.
By  selecting only  two  tracks identified  as an $\ee$ pair with energy $> 4$ GeV and invariant mass $M_{\ee} < 95~\rm MeV$, the inverse total momentum pointing back to the target, and   
 forming a vertex within the DC fiducial volume,  
an excess of 2.1 events compatible with background expectations was observed.
For a given flux $d\Phi(M_X, E_X, N_{pot})/d E_X$ of $X$'s the expected number of 
$\xdecay$ decays occuring within the fiducial length $L$ of the NOMAD 
detector located  at a distance $L'$ from the
neutrino target is given by 
\begin{eqnarray}
N_{\xdecay}=\brx Br(X\to\ee) \int \frac{d\Phi}{dE_X}\nonumber \\  
\cdot  exp\Bigl(-\frac{L'M_{X}}{P_{X}\tau_X}\Bigr)\Bigl[1-exp\Bigl(-\frac{L M_X}{P_X\tau_X}\Bigr)\Bigr]\varepsilon A dE_X
\label{flux}
\end{eqnarray}
where $ E_X, P_X$, and $\tau_X$ are the $X$ energy, momentum and the lifetime  
at rest, respectively,  $\varepsilon$ is the $\pair$ pair reconstruction efficiency,
$N_{pot}$ is the  number of pot.
The acceptance $A$ of the detector was calculated  tracing $X$'s
produced in the Be-target or beam dump to the detector taking the
relevant momentum and angular distributions into account. 
As an example for a mass $M_{X}= 10~\rm MeV$, $A=8.5\%$ and $\varepsilon\simeq 20\%$.\
The $X$ flux  calculated as a function of the $X$ energy is shown in Fig. \ref{spectr}.

The final 90$\%~C.L.$ exclusion region in the
$\brx Br(\xdecay)$ vs $\tau_X$ plane shown in Fig. \ref{limit} together with the PS191 result  (see below) is  calculated by using the relation $N_{\ee}^{90\%} > N_{\xdecay}$, where $N_{\ee}^{90\%}$ (= 2.1 events) is the 90$\%~C.L.$ upper limit for the expected number of signal events \cite{nomadnuh}. Our result is sensitive to a branching ratio $Br(\pi^{0}\to \gamma X) \gtrsim 10^{-15}$, which is more than nine orders of magnitude smaller than the previous limit from the  experiment\cite{mei1}. Over most of the $\tau_{X}$ region, the $X$  lifetime is sufficiently long, that $LM_{X}/p_{X}\tau_X \ll L'M_{X}/p_{X}\tau_X \ll 1$. 

\begin{figure}
\includegraphics[width=0.45\textwidth]{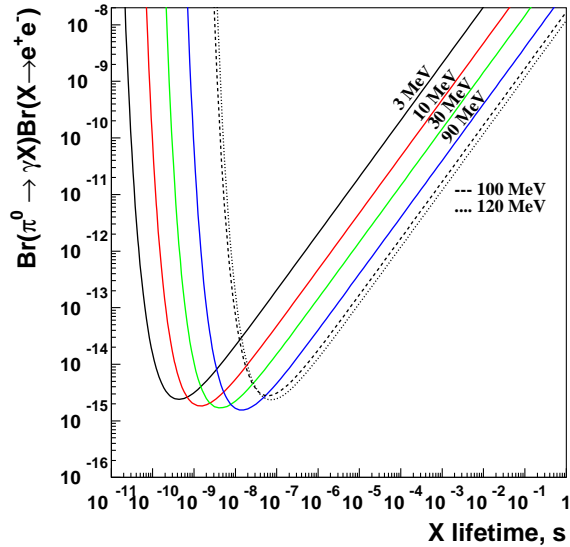}
\caption{ The 90$\%~C.L. $  upper limits on the branching ratio
$Br(\pi^{0}\to \gamma X) Br(X\to \ee)$ versus $\tau_{X}$ from the NOMAD (solid) and PS191 (dashed, dotted) experiments. The numbers  near the curves indicate the corresponding values of $M_X$. }
\label{limit}
\end{figure}  

To obtain  limits for the region $M_X> 95$  MeV, we consider next the  results of the PS191 neutrino experiments at CERN that searched for  decays $\nu_h \to \nu \ee$ of heavy neutrinos produced 
in 2-body  decays $\pi, K \to e,\mu+ \nu_h$ in the $\nu_h$ mass range from 
$\simeq 10$ to $\simeq 350$ MeV 
\cite{ps191-1, ps191-2}. This experiment, specifically 
designed to search for neutrino decays in a low-energy neutrino beam,  was performed 
by   using 19.2 GeV protons 
from the CERN Proton Synchrotron with the  total number of $0.86\times 10^{19}$ pot.
The PS191 detector, located at the distance of 128 m from the target,
 consist of a 12 m  long decay volume, eight chambers located 
inside the volume to detect charged tracks and  followed by a calorimeter.
The events searched for in the experiment were requested to consist of two tracks originating from a common vertex in the decay volume  and giving rise to at least one shower in the calorimeter.
No single $\ee$ events were observed and limits were established on the 
$\nu_{e,\mu} - \nu_h$ mixing strength as a function of the $\nu_h$ mass.
 
Using Eq.(\ref{flux}) for the PS191 case, similar to above considerations, 
we calculte the 90 $\% ~C.L.$  branching ratio limit curve shown in Fig. \ref{limit}
together with the result from NOMAD. For the  mass region
 $95 <M_{X}<$ 135 MeV/c$^{2}$   the best limits from PS191  are in the region $\brx Br(X\to\ee) \lesssim (2-4)\times 10^{-15}$.   In this estimate the  average $X$ momentum is
$<p_{X}>\simeq 1$ GeV and  the decay region length is $l=7$ m.
The results obtained were also cross checked with the limits  set by PS191 on the 
$\nu_{e,\mu} - \nu_h$ mixing strength \cite{ps191-1} and found to be in agreement.
For the mass region below  95 MeV, NOMAD provides better bounds than PS191.
The   $\pi^0 \to \gamma X$ decay rate constrained by NOMAD  is  small enough to produce a detectable excess of $\pair$ events in the PS191 experiment. In addition, for  masses $M_X \lesssim 10$ MeV the 
reconstruction efficiency drops, because the opening angle of the $\pair$ pair is $\simeq M_X/E_X \lesssim 10^{-2}$ rad, which  is small enough  to be resolved in the PS191 detector. Hence,
some events would be misidentified as a single track and, would be rejected.

The obtained results can be used to impose constraints on the magnitude of the coupling constant $\alpha_X = g_X^2 /4\pi$ for    the interaction of $X$ bosons with both quarks and leptons, which can be written in the form:
\begin{equation}
L_X= g_X (Q_{BX} B^i+Q_{eX} L_e^i+Q_{\mu X} L_\mu^i+Q_{\tau X} L_\tau^i) X^i
\end{equation} 
where $B^i = \sum_{q=u,d,s,..} \overline{q}\gamma^i q $, $L_e^i= \overline{e}\gamma^i e +\overline{\nu}_{eL}\gamma^i \nu_{eL}$, ..., see e.g. \cite{di, gkx2}.  Assuming charges $Q_{BX}\simeq Q_{eX} \simeq 1$,  we found 
\begin{equation}
\alpha_X < 3.4 \times 10^{-13}\frac{1}{M_X [MeV]} \Bigl(1- \frac{M_X^2}{M_\pi^2}\Bigr)^{-3/2}, 
\label{alphax}
\end{equation}
which is valid for $M_X < M_\pi$, where $M_\pi$ is the mass of $\pi^0$, and   
 $\tau_X \gtrsim 10^{-10} M_X$[MeV] s. This limit is more restrictive than those obtained in \cite{md, gkx2},   and than  bounds reported by NOMAD \cite{nomadx}. Less stringent limits (by a factor $\simeq \alpha$) could be obtained  for the cases where the $X$ interacts only with leptons, or when it is a leptophobic boson which interacts only with quarks  and  decays dominantly into $\ee$ pair through the quark loop   if its mass $M_X > 2m_e$ \cite{md, gkx2}.

We can also constrain recent models, where new  vector-like bosons mediate interaction between our world and hidden sectors due to the kinetic mixing term  
\begin{equation}
 L_{int}= -\frac{1}{2}\chi F_{\mu\nu} V^{\mu\nu}.
\label{mix}
\end{equation}
where,  $\chi$ is the mixing strength, and $F^{\mu\nu}$, $V^{\mu\nu}$ are the ordinary photon and hidden vector  field strengths, respectively, see e.g. \cite{prv}. As follows from \eqref{mix}, any source of 
$\gamma$'s could produce kinematically possible  massive states $V$  
according to the appropriate mixings. If the mass of $V$ is below   the mass of $\pi^0$, it can be produced in the  decay $\pi^0 \to \gamma X$, with the  subsequent decay of $X$ into $\ee$ pair \cite{bpr}. The corresponding branching ratio is given by: 
\begin{equation}
Br(\pi^0 \to \gamma V) = 2\chi^2\Bigl( 1- \frac{M_V^2}{M_\pi^2}\Bigr)^3
\label{br}
\end{equation}
\begin{figure}[htb!!]
\includegraphics[width=0.45\textwidth]{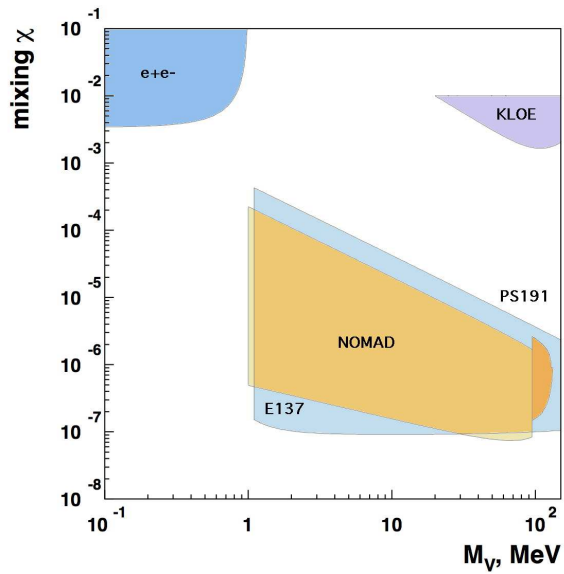}
\caption{ 90$\%~C.L.$ exclusion regions in the ($M_{V}; \chi $) plane 
obtained from the results of the NOMAD  \cite{nomadnuh}, PS191 \cite{ps191-1},  and positronium 
\cite{mitsui} experiments. The areas excluded by the electron beam dump 
experiment E137 \cite{e137} and by the 
KLOE experiment \cite{kloe} are also  shown for comparison.}
\label{coupl}
\end{figure}  
 For $V$  masses below the $\pi^0$-meson mass, 
$M_{V}\lesssim M_{\pi}$, the dominant $V$ boson decay is $\ee$  with a rate which,  
for small mixing,  is given by:
\begin{eqnarray}
\Gamma (\vdecay) = \frac{\alpha}{3} \chi^2 M_V \sqrt{1-\frac{4m_e^2}{M_V^2}} \Bigl( 1+ \frac{2m_e^2}{M_V^2}\Bigr)
\label{rate}
\end{eqnarray}
Using Eqs.(5,6) we can  determine the $90\%~ C.L.$ exclusion regions in the 
($M_{V}; \chi $) plane from the results of the NOMAD \cite{nomadnuh} and PS191 \cite{ps191-1} experiments, which are shown in Fig. \ref{coupl} together with the area excluded 
 by the electron beam dump experiment E137 at SLAC \cite{e137}, see 
 Ref.\cite{jdb} and discussion therein,   positronium experiment searching for the decay $\ee \to \gamma V$ \cite{mitsui} and recent
 results from KLOE \cite{kloe}. 
The shape of the exclusion contour  from the  PS191 experiment corresponding to the $X$ mass range
$M_V > 95$ MeV is defined mainly by the phase space factor in \eqref{br}.
One can see, that the NOMAD exclusion region falls
basically within that of the E137 experiment. However, the fact that 
the NOMAD events are generated by the proton beam, while the E137 events 
are originated from the electron beam dump is important if, e.g. the  $X$ is a
leptophobic boson which interacts dominantly with quarks.
The attenuation of the $X$- flux due to $X$ interactions with matter was found to be negligible, 
e.g. for couplings of \eqref{alphax} the $X$ boson mean free
path in iron is $\gg$ 100 km, as compared with the iron and earth shielding 
total length of 0.4 km used for the NOMAD beam.

 In summary, using sensitive limits from the PS191 and NOMAD experiments on heavy neutrino decays 
$\nu_h \to \nu \ee$ in the sub-GeV mass range, new stringent limits on the branching fraction  $Br(\pi^0 \to \gamma X) \lesssim 10^{-8}-10^{-15}$, for $X$ masses $1 < M_X < 135   $ MeV and 
  lifetime $10^{-11}< \tau_X < 1$ s, are obtained. 
Our best result is sensitive to a branching ratio 
$Br(\pi^{0}\to \gamma X) \gtrsim 10^{-15}$, which is about nine orders of magnitude
smaller than the previous limit from the SINDRUM experiment\cite{mei1}. It is also 
a factor $\simeq 100$ more stringent than the bound from cosmological considerations \cite{ng}.
The obtained results are used to set new limit on  the $X$ coupling strength
to lepton and quarks, and also to constrain models, in which mixing between photons and   a new vector-like bosons mediates interaction between our world and  
 hidden sectors consisting of  $SU(3)_C \times SU(2)_L \times U(1)_Y$ singlet fields.
 For the $X$ mass range $\lesssim M_\pi$  the obtained limits  on the mixing $\chi$ are more stronger than those recently derived by the experiment KLOE \cite{kloe},  and comparable 
 with those obtained in Ref.\cite{jdb, bur}. Stringent constraints on mixing strength $\eta$
 obtained from SM1987A observations, have been recently reported in \cite{dent}. 
   We demonstrate a significant  potential  of experiments similar to NOMAD for the future more sensitive searches for new physics at the high intensity frontier \cite{hif}. 
 For example,  stringent limits for the extended sub-GeV $X$ boson mass range $M_X > M_\pi$ could  
 be obtained from the searches for  $\eta, \eta'$ mesons decays  $\eta, \eta' \to \gamma X$ with the subsequent decays $X \to \ee, \mu^+\mu^-, \pi^+\pi^-,\mu^\pm e^\mp, \mu^\pm \pi^\mp,...$
We note that limits on the decays  $\eta, \eta' \to \gamma X$ could also be obtained from 
PS191 or other existing data provided  cross-sections for $\eta, \eta'$ productions in the forward 
direction in $p-Be$ collisions are known. These limits are expected to be  
more restrictive than those recently  reported by the  A1 collaboration from the search  for the decay $X\to \ee$ in a fixed target experiment  \cite{a1}.

I am indebted to N.V. Krasnikov for useful discussions, and M.M. Kirsanov for help 
in calculations.

\end{document}